\documentclass[eqsecnum,aps,epsfig]{revtex4}
\usepackage{amsmath,amssymb,amsthm,bm}
\usepackage{psfrag}
\usepackage[dvips]{graphicx}

\begin{document}
\title{Removing the concavity of the thick center vortex potentials by fluctuating the vortex profile}
\author{S.~Deldar$^\dagger$ and S.~Rafibakhsh$^\ddagger$}
\affiliation{
Department of Physics, University of Tehran, P.O. Box 14395/547, Tehran 1439955961,
Iran \\
$^\dagger$E-mail: sdeldar@ut.ac.ir\\
 $^\ddagger$E-mail: rafibakhsh@ut.ac.ir
}
\date{\today}

\begin{abstract}

The thick center vortex model reproduces important aspects of the
potentials between static quark sources as seen in lattice Yang-Mills
calculations: Both the intermediate distance behavior, governed by
Casimir scaling, as well as the long distance behavior, governed by
N-ality, are obtained. However, when a fixed vortex profile is used,
these two distance regimes do not connect naturally to each other. The
transition in general violates concavity constraints on the potential, especially for higher representations of the gauge group. We demonstrate how this issue
can be alleviated when the vortex profile is allowed to fluctuate within this
simple model.
\end{abstract}

\maketitle

\section{INTRODUCTION}\label{Intro}

The thick center vortex model is a phenomenological model which has been fairly successful in describing the mechanism of confinement in QCD \cite{Fabe1998}. One assumes that vortices are some special class of field configurations which fill out the QCD vacuum \cite{Hooft1979}. The potential between a pair of static quark-antiquarks is obtained by the interaction of the vortices with the world line of the static pair. It has been shown that confinement occurs due to random fluctuations in the number of vortices.

Experiments and theories suggest some features for the potentials between static quarks. Any model which aims to reproduce the potential should address these
features. The thick center vortex model provides the intermediate distance linear potentials \cite{Fabe1998,Deld2000,Deld2007} qualitatively in agreement with Casimir scaling. The large distance behavior which is governed by the $N$-ality of each representation, is also obtained successfully. The $N$-ality of each representation is given by mod($n-m$), where $n$ is the number of quarks and $m$ is the number of antiquarks constructing the representation. 
Recently, some works \cite{Fabe2008,Deld2008,Deld2009} have been done to reproduce the short distance Coulombic potential, as well. In addition, the intermediate distance linear potential has been studied and the extent of this regime has been increased \cite{Green2006}. However, the potentials show some unphysical concavity for intermediate regime where it is connected to the long distance regime. In this paper we discuss how to remove this artifact for intermediate distances. We should recall that there is also a convexity issue for the short distance regime which is not discussed in this paper.

In section two, some general features of the  potentials between static SU($N$) sources  are discussed. A brief review of the thick center vortex model is given in section 3. The method which leads to removing the concavity is presented in section 4. Chapter 5 concludes this article.

\section{Some general features of static quark potentials}

In general, there are three regions for the potential between static quark sources of the fundamental and higher representations. Quarks are the sources in the fundamental representation. Higher representation sources are constructed from different combinations of quarks and antiquarks. For convenient, we call all the sources as quarks throughout this article. In the following discussion, the behavior of the potentials in these regions is studied. 

a) At short distances a Coulombic behavior is expected. So the potential is proportional to $\frac{1}{R}$ for this region, where $R$ is the distance between the quark and antiquark pair. The Coulombic strength of each representation  should be proportional to the quadratic Casimir operator  of that representation. For this regime, perturbative methods work well, since the coupling constant is small enough and the potential between the source pair may be studied by the one gluon exchange procedure. However, one cannot describe  the potentials of other regimes by these methods.

b) For intermediate distances, one expects to see a linear potential proportional to $R$, in agreement with the lattice calculations. Based on lattice results \cite{Bali2000,Deld1999}, the coefficient of the linear part which is called the string tension, is proportional to the quadratic Casimir operator of the corresponding representation. This is called Casimir scaling. The Casimir scaling regime is expected to extend roughly from the onset of confinement to the onset of screening.

c) At large distances, where the distance between the quark-antiquark pair increases, the energy stored in the string is increased such that a pair of gluon-antigluons is created from the vacuum. The gluons combine with the initial sources and the new pair changes its dimension to the lowest dimension with the same N-ality. Thus, sources with the same $N$-ality obtain the same string tensions in this regime. Especially, those with zero $N$-ality are screened. Therefore, for example, for the SU($3$) gauge group, representations with zero $3$-ality are screened at large distances and others with $3$-ality equal to $1$ acquire the same string tension as the fundamental quarks. 

The static quark potentials must be convex every where\cite{Bachas1986} \footnote{The usage of ''convex" here (and in the related literatures) is at variance with the standard mathematical usage.  By convexity of the potential we mean that the potential is constrained to be increasing but concave downward according to equation \ref{convex}}. They should be monotonous non-decreasing and convex functions of $R$. This property is true, independent of the gauge group and the details of the matter sector. It means that:
\begin{equation}
\frac{dV}{dr}>0 \qquad ;\qquad \frac{d^2V}{dr^2} \leq 0
\label{convex}
\end{equation}
According to these inequalities, the force between a static quark-antiquark pair is always attractive but does not increase with distance. From the second inequality, one understands that the potential asymptotically can rise no faster than linearly with distance. Although, from these requirements, one cannot prove that the static quark potential rises linearly, however, lattice simulations and hadron phenomenology suggest that it is linear at medium and large quark separations. 

Of course, it would be very interesting if one finds a phenomenological model which describes the potentials for all regimes even though it is not compulsory. There has been some progress within the thick center vortex model using the trivial center element of the gauge group \cite{Fabe2008,Deld2008,Deld2009} which leads to a Coulombic behavior. In \cite{Fabe1998} it was shown that the Casimir scaling at the intermediate distances can be reproduced by the thick center vortices. The $N$-ality dependency at large distances is a well- known feature of the model, as well. However the concavity of the potentials which is a significant issue at the place where the two regimes connect to each other, is an open question. 

 In the next section, a very brief explanation of the thick center vortex model is represented and then in section four the concavity of the potential at intermediate distances and the possible solution are discussed.

\section{The center vortex model}

The center vortex model states that the QCD vacuum is filled by  some special line-like (in three dimensions) or surface-like (in four dimensions) objects, which carry a magnetic flux quantized in terms of the center elements of the gauge group \cite{Hooft1979}. When a vortex links to a Wilson loop, it derives a multiplicative factor $\exp({\frac{2\pi i n}{N}})\in Z_{n}$ where $\{n=1,2,...,N-1\}$. Using thin vortices, one gets the large distances potential between static sources proportional to the $N$-ality of the representation or $Z_{N}$ transformation. It means that for large distances one obtains a zero string tension for the adjoint and other zero $N$-ality representations and no intermediate distance linear potential in contrast to the lattice calculations which predict a linear regime even for zero $N$-ality representations. To obtain the intermediate potentials in agreement with the lattice data, vortices have to have a finite thickness \cite{Fabe1998}.

Using thick vortices \cite{Fabe1998}, the factor $\exp({\frac{2\pi i n}{N}})$ is replaced by a group factor $G(x,s)=S\exp[i\vec{\alpha}^{n}_{C}(x).\vec{H}]S^{\dag}$ where $\{H_{i},i=1,2,...,N-1\}$  are the generators spanning the Cartan sub-algebra and $S$ is an element of the SU($N$) gauge group. The vortex profile is given by $\alpha_{C}(x)$ which depends on the fraction of the vortex core enclosed by the loop $C$. Thus, if a vortex of type $n$ is located completely inside the loop, $\alpha_{C}(x)$ obtains its maximum value and if it is not linked to the loop it would be equal to zero. Otherwise, if the vortex links to the perimeter of the loop, the percentage of the linking between the vortex and the Wilson loop is calculated by the flux $\alpha_{C}(x)$. Based on this idea, the potential between the two static sources in the representation $r$ of the $SU(N)$ gauge group is given by:
\begin{equation}
V_{r}(R) = -\sum_{x}\ln\{ 1 - \sum^{N-1}_{n=1} f_{n}
            (1 - \mathrm{Re} {\cal G}_{r} [\vec{\alpha}^n_{C}(x)])\}
\label{sigmac}
\end{equation}
where  $x$ is the location of the center of the vortex and $n$ represents the vortex type. $f_{n}$ is the probability that any given unit area is pierced by a vortex type $n$.
\begin{equation}
{\cal G}_{r}[\vec{\alpha}] = \frac{1}{d_{r}} \mathrm{Tr} \exp[i\vec{\alpha} . \vec{H}]
\label{gr1}
\end{equation}
$d_{r}$ is the dimension of the representation.
Using this model, the intermediate distance string tensions exhibit
Casimir scaling and the long distance potentials are governed by N-ality,
in agreement with lattice calculations. However, for intermediate
distances, what is in agreement with the lattice calculations is only the Casimir scaling not the functional dependence on the separation throughout the whole area of Casimir scaling regime which is extended from the onset of confinement to the onset of screening. Especially, at the place where the potential is reaching to the onset of screening, some kind of unexpected concavity is observed for several representations which explicitly disagrees with lattice calculations where a monotonous non-decreasing and convex functions of $R$ is obtained. 

\begin{figure}[]
\begin{center}
\vspace{70pt}
\resizebox{0.47\textwidth}{!}{
\includegraphics{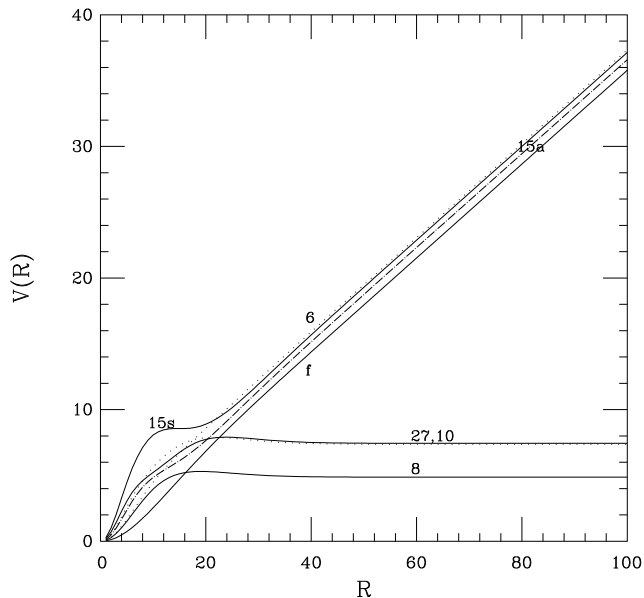}}
\vspace{-20pt}
\caption{\label{origin}
Potentials between static sources of SU($3$) gauge group using thick center vortex model. Although the  large distance potentials is obtained in agreement with the theoretical and experimental evidences, and the intermediate string tensions are qualitatively in agreement with Casimir scaling but the observed concavity of the potentials for $10 \leq R \leq 20$ does not seem to be physical and disagrees the lattice results.}
\end{center}
\end{figure}

Figure \ref{origin} shows the potentials between static sources for a typical flux distribution for several representations of the $SU(3)$ gauge group obtained from the thick center vortex model. The concavity is observed, especially very clearly for representation $15_{s}$ for $10\leq R \leq 20$. Even though it is possible to reduce this concavity by using alternative fluxes, it is not possible to remove it completely this way.

In the next section, we show that this problem may be solved by fluctuating the vortex profile. We would like to mention again that there is also a convexity issue at short distances which is not addressed by this paper and we only discuss the intermediate distance regime. 

\section{Fluctuating the vortex profile}

In general, any vortex profile which satisfies the following conditions and leads to a physical potential is acceptable:

1. vortices which pierce the plane far outside the Wilson loop do not affect the loop.
In other words, for fixed $R$, as $x\rightarrow \infty$,
$\alpha \rightarrow 0$.

2. If the vortex core is entirely contained within the Wilson loop, then ${\cal G}_{r}[\vec{\alpha}]$ would be equal to one of the center elements and $\alpha$ would get the maximum value.

3. As $R\rightarrow 0$ then $\alpha \rightarrow 0$.

A physical potential is a potential which shows a linear behavior at intermediate distances and an asymptotic behavior proportional to the $Z_{N}$ transformation of the representation at large distances. In principle, it may be possible to obtain  the best profile from the lattice calculations but for sure it is a very hard job. Another approach is to find the best profile by examining different ones and finding those which agree better the lattice results. Of course, within the simple model like the thick center vortex model which assumes that the probabilities, $f_{n}$, that vortices pierce a given area are not correlated, this might not be easily achieved. A variety of fluxes has been studied in reference \cite{Deld2000}. The results show that by using axially symmetric distributions, one gets the string tensions which are qualitatively in agreement with the Casimir scaling. However, this agreement is lost if one uses non-axially symmetric distributions. There are lots of profiles which give the physical potentials we have discussed above. 

In general, using the thick center vortex model, one chooses a fixed vortex profile and applies it uniformly through the space-time and then $V(R)$ is calculated from equation (\ref{sigmac}) for each $R$. Now we assume some fluctuation for the profile and  calculate the potential from equation  (\ref{sigmac}) several times, each time with a random flux which is slightly different from the previous one. At the end, we obtain the potential by averaging the potentials obtained from different fluxes. In this method, once the profile is chosen, the same profile is used for the whole space-time.

\begin{figure}[]
\begin{center}
\vspace{70pt}
\resizebox{0.47\textwidth}{!}{
\includegraphics{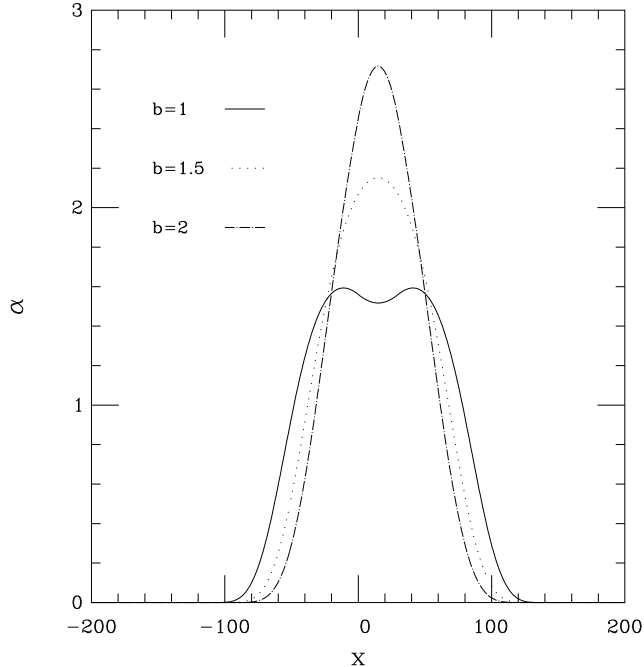}}
\vspace{-20pt}
\caption{\label{flux}
Profile function $\alpha_R(x)$ of Eq.~(\ref{alphar}) with $a=150$ and different $b$'s for $R=30$.}
\end{center}
\end{figure}

To practice this idea, one has to choose a flux with the capability of being changed easily with the parameters of the model. Changing the profile should not result to an unphysical potential such that the Casimir scaling at intermediate region or the  dependence to  the $N$-ality for the large distances are lost.
Using the following flux, we have been able to obtain a variety of physical fluxes by changing one of the parameters.
If we put the legs of the Wilson loop at $x=0$ and $x=R$, then $\alpha_{c}(x)$, the fraction of the flux within the loop, is given by:
\begin{equation}
\alpha_{c}(x)=\beta(x)-\beta(x-R)
\label{alphar}
\end{equation}
Where $\beta(x)$, the amount of the vortex flux contained in different regions is chosen as the following:
\begin{equation}
\beta(x)= \left \{ \begin{array}{llll} 
\frac{2\pi}{\sqrt{3}}&\quad x\ge a\\
\frac{2\pi}{\sqrt{3}}(1-\exp[b(1-\frac{1}{({\frac{x}{a}-1)}^{2}})])  &\quad 0<x<a\\
-\frac{2\pi}{\sqrt{3}}(1-\exp[b(1-\frac{1}{({\frac{x}{a}+1)}^{2}})])  &\quad -a<x<0\\
-\frac{2\pi}{\sqrt{3}}&\quad x\le -a\\
\label{beta-flux}
\end{array}
\right.
\end{equation}
$a$ and $b$ are free parameters of the model. To satisfy the three conditions of the beginning of this section, $\beta(x)$ must be equal to $\pm \frac {1}{2}\alpha_{max}$ when $x\rightarrow \pm \infty$, where $\alpha_{max}$ is the maximum value of $\alpha_{c}(x)$.

By slightly changing the free parameter $b$, the flux is fluctuated.
Figure \ref{flux} shows some of the fluxes that may be obtained from this flux by changing the parameter $b$. For this plot, the parameter $a$ is assumed to be fixed and equal to $150$ and the flux is plotted for $b=1$,$1.5$ and $2$. As figure \ref{flux} indicates, for $b=1$ the flux shape shows a minimum at the center of the vortex. As $b$ decreases to the values less than $1$, this valley at the center of the vortex becomes deeper. As a result, the potentials obtained for $b<0.5$ show some unphysical behavior like the Casimir scaling violation as discussed in \cite{Deld2000}. Therefore, choosing $a=150$, the lowest value for $b$ is chosen to be $0.5$. The upper limit of $b$ is chosen such that the potentials for all the representations we are studying, show a convex behavior. As seen below, an upper limit of b=10 for the fluctuating vortex
profile is sufficient to remove the concavity of the potentials for all
representations, especially $15_{s}$ which has shown the worst concavity as shown in figure \ref{origin}. Of course the value of $b$ is not that much rigid, especially for the upper limit. One may vary it more or less as far as the shape and the scale of the potential is not changed drastically for the interval chosen for the parameter $b$. It is observed that by this method, different flux shapes are obtained. However, we have not used non-axially symmetric distributions which destroy the ordering of the potentials compared with the Casimir scaling. we should recall that although not for any choice of parameters but in a large region of the parameters space of $a$ and $b$, the intermediate Casimir scaling and large distance $N$-ality dependence are obtained. Even though the possibilities are infinite but some parameters like what we have chosen, work better than others.

\begin{figure}[]
\begin{center}
\vspace{70pt}
\resizebox{0.47\textwidth}{!}{
\includegraphics{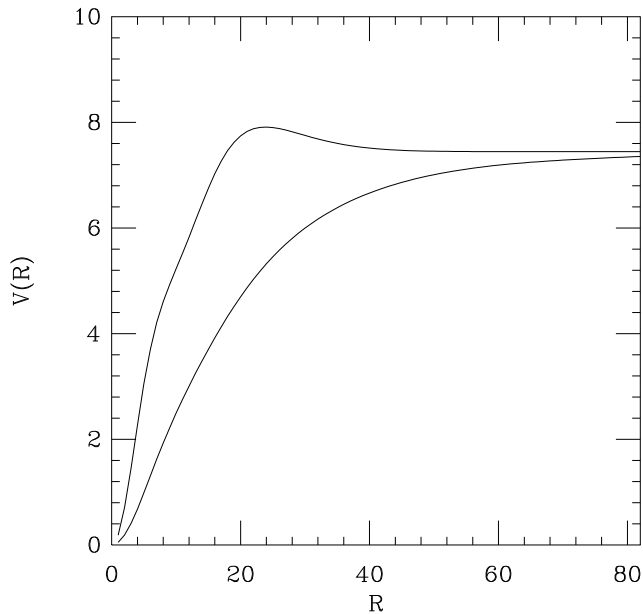}}
\vspace{-20pt}
\caption{\label{both27}
Potential between two static sources of representation $27$ of SU$(3)$ gauge group. For the upper plot, we have used one of the profiles of  Eqs. ~(\ref{alphar}) and ~(\ref{beta-flux}) with $a=150$ and $b=10$. For the lower plot, we have used fluxes generated from the same equation with the same $a$ but but averaged over various $b$ between $0.5$ and $10$.}
\end{center}
\end{figure}

\begin{figure}[]
\begin{center}
\vspace{70pt}
\resizebox{0.47\textwidth}{!}{
\includegraphics{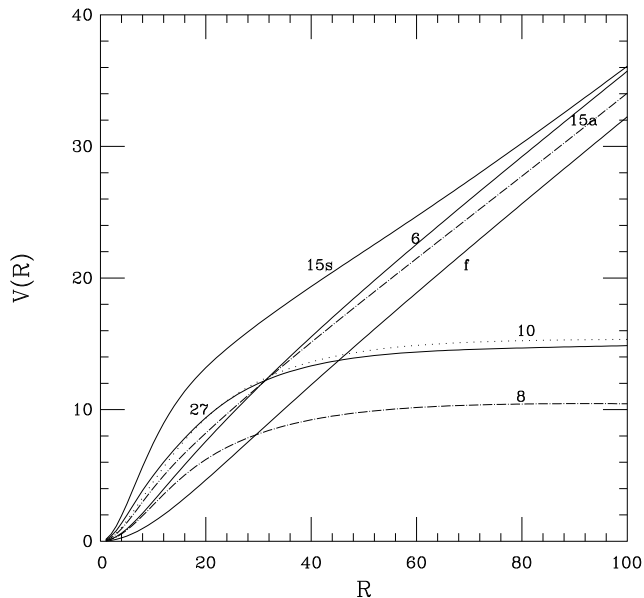}}
\vspace{-20pt}
\caption{\label{all}
Potentials between static sources of SU($3$) gauge group using averaged fluctuated fluxes which is generated from the profile of equations (\ref{alphar}) and (\ref{beta-flux}). $a=150$ and $b$ is chosen randomly between $0.5$ and $10$. Compared with figure \ref{origin}, the unphysical concavity is completely removed for intermediate distances.}
\end{center}
\end{figure}

To obtain the potential, we have used several fluxes with $a=150$  and $b$ which has been chosen randomly between $0.5$ and $10$. There is no weighting for different $b$'s and all the values are used equally. Figure \ref{both27} shows the plots of the potentials between two sources of representation $27$.  The upper plot represents the potential using only one flux with $a=150$ and $b=10$. The lower plot shows the averaged potential using fluctuated fluxes. $a$ is equal to $150$ and $b$ changes randomly between $0.5$ and $10$. It is seen that the concavity is gone for the lower plot. This method has been tested for a variety of sources in different representations of $SU(3)$ and $SU(4)$ gauge groups and the results are satisfactory. The potentials at intermediate distances are still proportional to the Casimir scaling and for large distances the slopes of the potentials are in agreement with the $N$-ality. The potentials of zero $N$-ality representations are screened and representations with the same $N$-ality obtain the same slope. Figure \ref{all} shows the potential between various SU($3$) sources using the fluctuated flux. Compared with figure \ref{origin}, it is observed that the concavity of the intermediate distances is completely removed for all representations. For this plot, $1000$ fluxes with random $b$ between $0.5$ and $10$ are used and at the end the potential is averaged. For figure \ref{origin}, we have used equation (\ref{alphar}) and (\ref{beta-flux}) with $a=150$ and $b=10$.

It seems reasonable to fluctuate the vortices since what is happening in this model is that the full Yang-Mills path integral is truncated to a subset of vortex-type configurations, and it seems
perfectly natural to include in this subset a whole range of vortex profiles. Vortices are not classical solutions of Yang-Mills theory, and there is therefore no argument which would compel one to extend the truncated path integral only over configurations of one particular profile; on the contrary, the converse seems more natural.  However, the mechanism of changing the profile must be much more complicated than random procedure, we have assumed. But, it seems very implausible to enter those complicated assumptions into such simple model, the thick center vortex model.

\section{Conclusions}

Even though, the thick center vortex model has been able to reproduce aspects of the potential between static sources for large and intermediate distances, there are still some shortcomings within the model which disagree the theories and experiments. Both the Casimir scaling behavior at intermediate distances and the
N-ality dependence at large distances have been generated correctly;
however, the potentials show concave behavior at the place where the intermediate regime is connected to the large distances regime. This concavity is very clear for some representations. In this paper, we have shown that by fluctuating the vortex profile and then averaging the potentials obtained from those fluxes, the concavity would be removed. 

\section{\boldmath Acknowledgments}

We would like to thank Manfried Faber and Denise Neudecker for the very helpful discussions. We are grateful to the research council of University of Tehran for supporting this study.

\end{document}